\UseRawInputEncoding
\documentclass[reprint,
superscriptaddress,
amsmath,amssymb,
aps,prl,prb,rmp,prstab,prstper,floatfix]{revtex4-2}
\usepackage{graphicx}  
\usepackage{dcolumn}   
\usepackage{bm}        
\usepackage{verbatim}   
\usepackage{placeins}
\usepackage{amsmath,amssymb}
\usepackage{textcomp} 
\usepackage{enumitem}
\usepackage{color}
\usepackage{psfrag}
\usepackage[colorlinks,citecolor=blue,linkcolor=blue,urlcolor=blue]{hyperref}
\usepackage{subfigure}
\usepackage{amsfonts}
\usepackage{epstopdf}
\usepackage{titlesec}
\usepackage{balance}
\usepackage{flushend}
\begin{document}
	
\title{Verified Universal Breakdown of Kibble-Zurek Scaling in Fast Quenches}
	
\author{Xinxin Rao}
\thanks{These authors contributed equally to this work.}
\affiliation{School of Physics and Astronomy, Sun Yat-Sen University, Zhuhai, 519082, China}
	
\author{Yang Liu}
\thanks{These authors contributed equally to this work}
\email{\protect\\
	liuyang@quantumsc.cn}
\affiliation{Quantum Science Center of Guangdong-Hong Kong-Macao Greater Bay Area, Shenzhen 518045, China}
\affiliation{Guangdong Provincial  Key Laboratory of Quantum Metrology and Sensing, Sun Yat-Sen University, Zhuhai, 519082, China}

\author{Mingshen Li}
\affiliation{School of Science, Sun Yat-sen University, Shenzhen 518100, China}

\author{Teng Liu}
\affiliation{School of Physics and Astronomy, Sun Yat-Sen University, Zhuhai, 519082, China}

\author{Huabi Zeng}
\affiliation{School of Physics and Optoelectronic Engineering,  Hainan University, Haikou, 570228, China}

\author{Le Luo}
\email{luole5@mail.sysu.edu.cn}
\affiliation{School of Physics and Astronomy, Sun Yat-Sen University, Zhuhai, 519082, China}
\affiliation{Quantum Science Center of Guangdong-Hong Kong-Macao Greater Bay Area, Shenzhen 518045, China}
\affiliation{Guangdong Provincial  Key Laboratory of Quantum Metrology and Sensing, Sun Yat-Sen University, Zhuhai, 519082, China}
\affiliation{Shenzhen Research Institute of Sun Yat-Sen University, Shenzhen 518057, China}
\affiliation{State Key Laboratory of Optoelectronic Materials and Technologies, Sun Yat-Sen University, Guangzhou 510275, China}

\date{\today}
\begin{abstract}
The Kibble-Zurek mechanism (KZM) predicts that when a system is driven through a continuous phase transition, the density of topological defects scales universally with the quench rate. Recent theoretical work [H.-B. Zeng \textit{et al.}, \textit{Phys. Rev. Lett.} \textbf{130}, 060402 (2023)] has challenged this picture, showing that under sufficiently fast quenches, both the defect density and freezing time become independent of the quench rate and instead scale universally with the quench range. Here, we experimentally test this prediction using a single trapped-ion qubit to simulate fast quantum quenches in the Landau-Zener and 1D Rice-Mele models. We identify a critical quench rate \( v_c \) that scales with the quench range \( \delta_{\max} \), separating two distinct dynamical regimes. In the Rice-Mele model, for \( v < v_c \), the defect density follows the KZM scaling \( \sim v^{1/2} \); for \( v > v_c \), it exhibits a universal scaling \( \sim \delta_{\max} \), independent of the quench rate. Our results provide direct experimental evidence of the predicted breakdown of KZM universality under fast quenches.	
\end{abstract}

\maketitle

When a system is driven through a continuous phase transition, the spontaneous formation of topological defects is well described by the celebrated Kibble-Zurek mechanism (KZM) \cite{zurek1985cosmological,zurek1991decoherence,kibble2007phase}. The KZM predicts that the density of defects \( n \) scales with the inverse of the quench time \( \tau \) as $n \sim \tau^{-d\nu/(z\nu + 1)}$, where \( d \) is the spatial dimensionality of the system, \( \nu \) and \( z \) is the correlation length critical exponent and the dynamic critical exponent, respectively\cite{laguna1997density}. This scaling has  been extensively validated in a wide variety of physical systems\cite{chaikin1995principles}. 
However, the breakdown of this scaling occurs when critical assumptions fail, such as the absence of a critical point\cite{chaikin1995principles}, the breakdown of the adiabatic-impulse approximation\cite{dziarmaga2010dynamics}, finite-size effects\cite{xia2020winding,del2010structural,corman2014quench}, strong many-body interactions \cite{huang2021observation,liu2021dynamic}, significant coupling to an external environment\cite{dutta2016anti,weinberg2020scaling,king2022coherent}, among other factors.

Recently, the breakdown of Kibble-Zurek (KZ) scaling has been observed in ultracold atomic systems subjected to fast quenches across phase transitions~\cite{donadello2016creation,ko2019kibble,goo2021defect,goo2022universal,liu2018dynamical}. Experimentally, a plateau in the defect density was reported at fast quench rates, which has been attributed to defect saturation or early coarsening. These observations have raised considerable interest in the breakdown of KZ scaling under fast quenches~\cite{chesler2015defect,sonner2015universal,gomez2019universal}, where a quantitative understanding of the dynamics remains elusive. In particular, the underlying mechanism responsible for the emergence of the defect density plateau is not fully understood, the critical quench rates at which KZ scaling fails have not been precisely identified, and the dependence of the plateau value on the quench range remains to be established. Moreover, whether such behavior is universal across different physical systems remains an open question. To address these fundamental issues, Zeng \textit{et al.} have recently proposed a universal mechanism for the breakdown of KZ scaling in the fast-quench regime~\cite{zeng2023universal}.

\begin{figure}[htbp]
	\centering
	\includegraphics[width=0.45\textwidth]{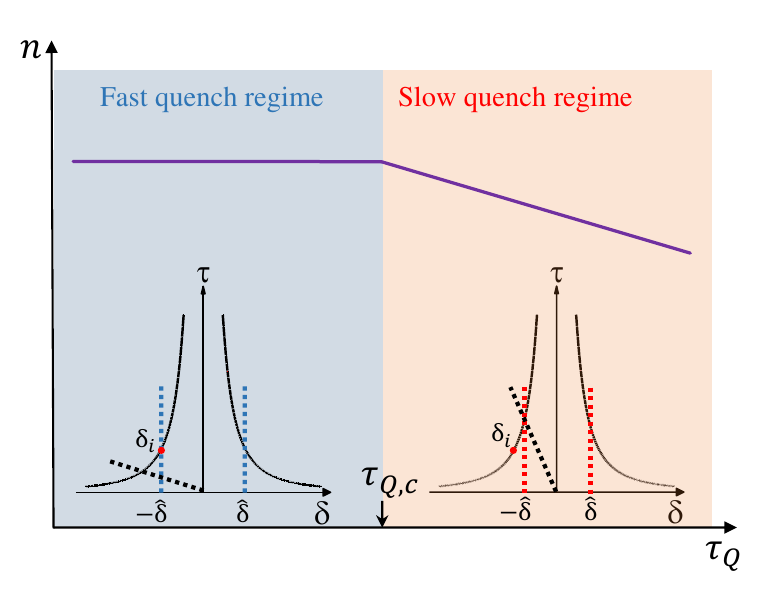}   
	\caption{Universal breakdown of the KZ scaling under fast quenches. The purple solid line represents the dependence of the defect density $n$ on the inverse quench rate $\tau_Q$. Left panel: $\tau_Q<\tau_{Q,c}$ the fast quench regime. The intersection between the inverse quench rate $\tau_Q=\tau/\delta$ (the dashed black line) and the relaxation time (the solid black line) lies outside the quench range $\delta_i$. As a result, $\delta_i$ determines the freezing region $\hat{\delta}$ (the dashed blue line), and the defect density $n$ saturates. Right panel: $\tau_Q>\tau_{Q,c}$ the slow quench regime. The intersection falls within $\delta_i$. Thus, the freezing region $\hat{\delta}$ (the dashed red line) depends on the $\tau_Q$, following the conventional KZ scaling.}
	\label{setup}
\end{figure}

This universal breakdown mechanism is illustrated schematically in Fig.~\ref{setup}. Within the conventional Kibble-Zurek framework, shown in the right panel, the solid black curve represents the divergence of the relaxation time near the critical point, while the dashed line denotes the inverse quench rate. Their intersection defines the freezing time, marking the boundary of the impulse region where the system's dynamics effectively halt. The relaxation time at this point sets the correlation length, which determines the defect density. Accordingly, KZM predicts a universal power-law scaling of defect density with the quench rate. Importantly, when the freezing point coincides with the boundary of the quench range, the corresponding quench rate defines a critical value that delineates the onset of the universal breakdown regime, named as the critical quench rate $v_c$. As shown in the left panel of Fig.~\ref{setup}, when the quench rate exceeds $v_c$, the intersection between the inverse quench rate and the relaxation time falls outside the quench range. In this regime, the relaxation time is limited by the quench range rather than the rate. Consequently, the correlation length---and thus the defect density---saturates, becoming independent of the quench rate and governed solely by the quench range. Both the saturated defect density and the critical quench rate exhibit universal power-law scaling with the quench range.

Given the fundamental implications of the predicted universal breakdown of the KZM, experimental verification is imperative. However, traditional platforms, such as solid-state materials or ultracold atomic gases, face technical barriers in reaching the fast-quench regime, where the quench rate significantly exceeds the critical value. For example, in cold atom experiments, the quench rate is often constrained by hysteresis in magnetic coils used to tune interaction parameters.

To circumvent these limitations, we employ a trapped-ion system, which allows precise control over quench dynamics and avoids such technical constraints. We first realize the Landau-Zener (LZ) model, a canonical system for continuous phase transitions~\cite{damski2005simplest,zurek2005dynamics,damski2006adiabatic}. In this model, the inverse minimum energy gap emulates the divergence of the relaxation time near a critical point~\cite{xu2014quantum,cui2016experimental}. Our results reveal clear breakdown of KZM scaling: both the critical quench rate and saturated defect density follow distinct power-law dependence on the quench range, in agreement with theory. To further test the universality of the scaling behavior, we implement the one-dimensional Rice-Mele model by encoding the Bloch Hamiltonian into the qubit energy space, where quasimomentum is mapped to coupling strength~\cite{dora2019kibble,xiao2021non}. Again, we observe the power-law scaling with the quench range, where the exponents of the power laws are determined by $d,z,\nu$, directly confirming the universal features predicted in Ref.~\cite{zeng2023universal}.

\begin{figure}
	\centering
	\includegraphics[width=0.4\textwidth]{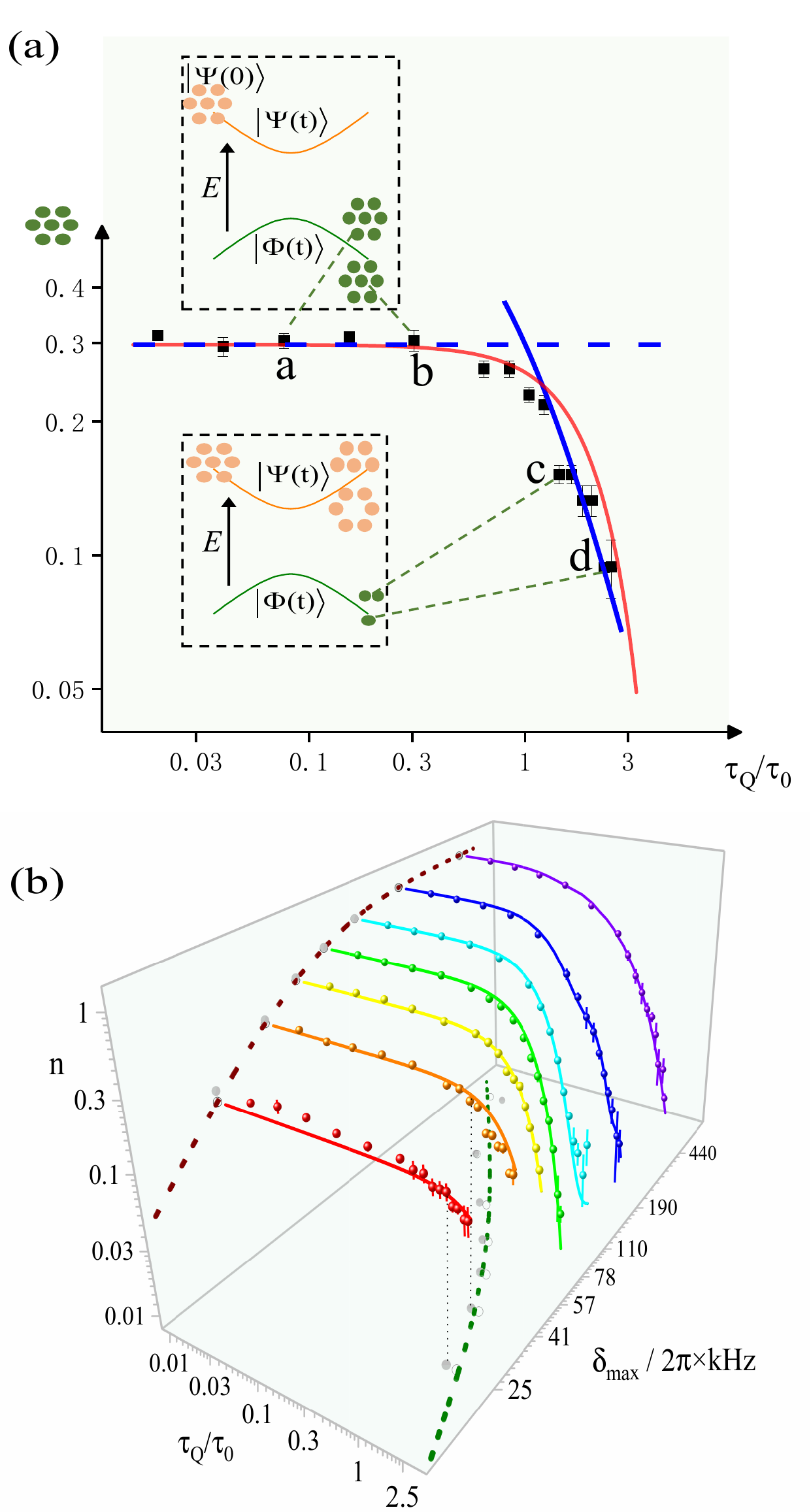}   
	\caption{Universal breakdown of the KZ scaling in the LZ model. (a) The dependence of the defect density $n$ on the dimensionless quench time $\tau_Q/\tau_0$ under a finite quench range. The upper and lower inset shows the cartoon picture for the phase transition, where the population of $\Psi(0)$ (the initial state) and $\Phi(T)$ (the defect density) are represented by number of orange and green circles, respectively. The upper (lower) inset shows the fast (slow) quench regime, where the points $a$ and $b$ ($c$ and $d$) with the different quench time have the same (different) defect density that breaks (follows) the KZ scaling. Black squares are experimental data at $\delta_{max}=2\pi\times 41$kHz. The solid red (simulation by the LZ model), dashed blue (prediction by the universal KZ breakdown) and solid blue (fitting to the KZ scaling) lines  are detailed in Supplemental Material S2. (b)The dependence of $n$ on both the quench range $\delta_{max}$ and $\tau_Q/\tau_0$. The solid red (orange, yellow, green, cyan, blue, purple) spheres and their corresponding lines are the experimental data and the simulation from LZ model for $\delta_{max}=2\pi \times 25$ ($41$, $57$, $78$, $110$, $190$, $440$) kHz, respectively. In the side plane, the gray solid circles (open circles and the dashed brown line) represent the saturated defect density of the experiment data (prediction by the universal KZ breakdown). In the bottom plane, the gray solid circles (the dashed green line, open circles) represent $\tau_{Q,c}/\tau_0$ obtained from the experiment data (fitting to universal KZ breakdown, prediction by the LZ model).}
	\label{defectdensity1}
\end{figure}

\emph{Experiment with the LZ model.} The experimental setup has been used in our previous works~\cite{lu2025dynamical,liu2025chiral,song2024non,lu2024realizing,lu2024experimental,bian2023quantum} and described in detail in Supplementary Material S1. A single $^{171}Yb^+$ ion has qubit states, which are the first-order magnetic field-insensitive hyperfine "clock" states of the $^{2}S_{1/2}$ level, $|0\rangle=|F = 0, m_F = 0\rangle$ and $|1\rangle=|F = 1, m_F= 0\rangle$, separated by 12.6 GHz. The qubit is initialized to $|0 \rangle$ by optical pumping. Then, the Landau-Zener tunnelling\cite{damski2005simplest,zurek2005dynamics,damski2006adiabatic} is realized by driving the qubit across the minimum of an energy gap, as shown in Fig.~\ref{defectdensity1} (a). 

The evolution of single qubit follows a time-dependent Hamiltonian
\begin{equation}
H(t)=J\cdot\sigma_x-\delta(t)/2\cdot\sigma_z
\end{equation}
where $J$ and $\delta$ is the strength and frequency detuning of the driving microwave, respectively. The eigenvalues of the Hamiltonian are $\lambda(t)=\pm\sqrt{J^2+(\delta(t)/2)^2}$. The microwave detuning is linearly modulated by $\delta(t)=-\delta_{max}+4\delta_{max}\cdot t/T$ with a constant $J$ and the quench range $\{-\delta_{max},\delta_{max}\}$ in the quench time of $T/2$. Initially, the qubit is prepared to the upper eigenstate of $H(t=0)$ with $|\Psi(t=0) \rangle = \{\lambda(0)+\delta_{max}/2, J\}^T $, while the lower eigenstate $|\Phi(t=0) \rangle = \{-\lambda(0)+\delta_{max}/2, J\}^T $. During the evolution, the qubit would first evolve adiabatically and stay in the upper instantaneous eigenstate. Near $t=T/4$, the qubit arrives in the vicinity of the avoided level crossing, i.e. $\delta(T/4)=0$, where the energy gap between two instantaneous eigenstates reaches the minimum value $2J $, and the tunnelling from the $|\Psi(t) \rangle$ to $|\Phi(t) \rangle$ would occur. This indicates the onset of the phase transition where the system enters the impulse region \cite{damski2005simplest}. As $\delta$ continues to vary, the system leaves the impulse region and returns back to the adiabatic region. 

We test the scaling laws associated with the breakdown of KZM. We assume that the qubit reach the equilibrium with the final state at $t_{f}=T/2$ (the assumption is applied throughout the study), and measure the density matrix of the system by quantum state tomography. The resulted defect density can thus be estimated by $n=|\langle \chi(t_{f})|\Phi(t_{f})\rangle|^{2}=Tr(\rho_{|\chi(t_{f})\rangle}\rho_{|\Phi(t_{f})\rangle})$, whose details are shown in Supplementary Material S2. By varying the period $T$ of this process, we obtain the dependence of the defect densities on the inverse quench rate, shown in Fig.\ref{defectdensity1} (a). A plateau of the defect density emerges when the dimensionless quench time $\tau_Q/\tau_0$ is small with $\tau_Q=2J/\dot{\delta}(t)$ and $\tau_0=1/2J$. The coupling $J=2\pi \times 31.75$ kHz throughout the LZ experiments. 

By repeating the above experiments for variable $\delta_{max}$, we obtain the dependence of the defect density on both $\tau_Q$ and $\delta_{max}$, as shown in Fig.\ref{defectdensity1} (b). Here, we observe two features. First, the defect density of the plateau, in the KZM-breakdown regime, is approaching the unity for increasing $\delta_{max}/2J$. Second, the critical inverse quench rate  $\tau_Q$ increases   with increasing $\delta_{max}/2J$. This indicates that the critical quench time correlates strongly with the quench range. These two features can be well explained by the universal breakdown mechanism described below.   

In the LZ tunneling system, the impulse region is set by $\hat{\delta}=\delta_{max}$ at a critical quench rate $v_c$, then the critical relaxation time can be described as $\tau_{c}=\tau_0/\sqrt{1+\epsilon_{c}^{2}}$, where $\tau_0=1/2J$ and $\epsilon_{c}=\delta_{max}/2J$\cite{damski2005simplest}. Correspondingly, the critical freezing timescale can be written as $\hat{t}_{c}=\tau_{c}/\alpha=1/(\alpha \sqrt{4J^2+\delta^{2}_{max}})$ and $v_c=\delta_{max}/\hat{t}_{c}=\alpha \delta_{max}\sqrt{4J^2+\delta^{2}_{max}}$. Consequently, the dimensionless critical quench time can be expressed by
$\tau_{Q,c}/\tau_0=J^{2}T/\delta_{max}=4J^{2}/v_c\propto 1/(x_c\sqrt{1+x^2_c})$, where $x_c=\delta_{max}/2J$. This equation is used for the fitting of the critical dimensionless quench time, as shown by the dashed green line in Fig.~\ref{defectdensity1}(b). When the quench rate $v \ll v_c$, the scale of the impulse region $\hat{\delta}=v*\hat{t}_{kz}\ll \delta_{max}$ with $\hat{t}_{kz}=(-2J^{2}/v^{2}+\sqrt{4J^{4}\alpha^{4}+v^{2}\alpha^{2}}/v^{2}\alpha^{2})^{1/2}$\cite{damski2005simplest}, representing the start of the quench lying outside the impulse region. In this case, the defect density follows the scaling $n(t)=\epsilon(t)^{2}/(1+\epsilon(t)^{2})=\delta(t)^2/(4J^2+\delta(t)^2)$\cite{damski2005simplest}, which is consistent with the adiabatic-impulse approximation. In contrast, when the quench rate $v\gg v_c$, the whole quenching process occurs inside the impulse region, and the effective impulse region is governed by $[-\delta_{max},\delta_{max}]$ with the freezing time $\hat{t}_c$, so the defect density becomes a constant and can be described by $n=|\langle \chi(t_f=T/2)|\Phi(t_f)\rangle|^2\approx |\langle \chi(t=0)|\Phi(t_f)\rangle|^2=\epsilon^{2}_i/(1+\epsilon^{2}_i)=x^2_{c}/(1+x^2_{c})$,
where $|\chi(t)\rangle$ is the instantaneous state of the qubit and could be derived analytically. This calculated $n$ is shown as the dashed brown line in Fig.~\ref{defectdensity1}(b). The detailed derivation of $|\chi(t)\rangle$ and $n$ are shown in the Supplemental Material S2\cite{supplementary}. 

\begin{figure*}[htb]
	\centering
	\includegraphics[width=1.0\textwidth]{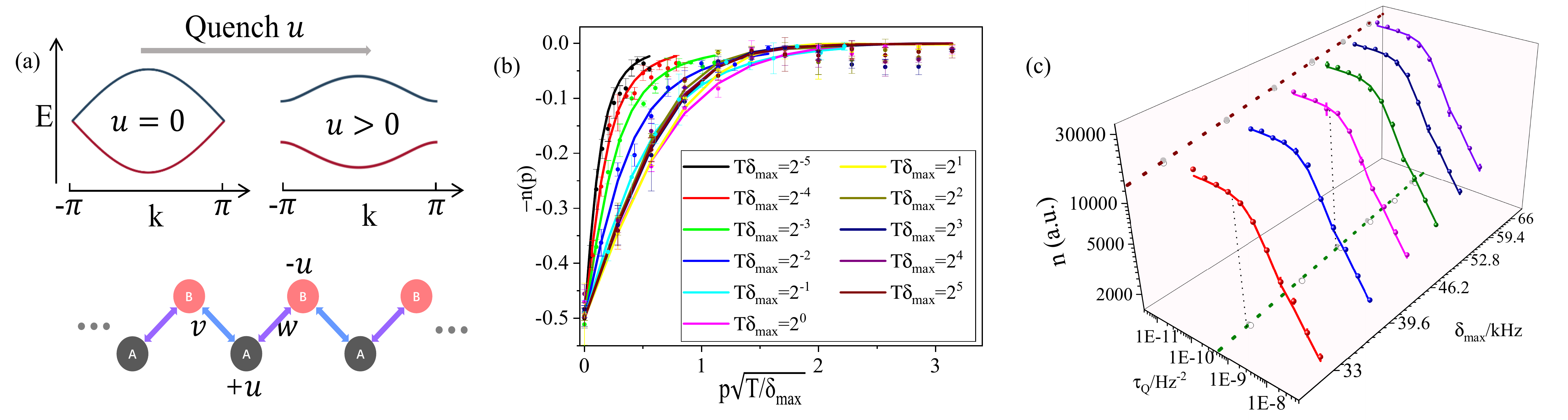}  
	\caption{Universal breakdown of the KZ scaling in the 1D Rice-Mele model. (a)The schematic of the Rice-Mele model, featuring intercell and intracell coupling $v$ and $w$ as well as the staggered onsite potential $u$ that is quenched from zero to a finite value, driving the transition from the metal to insulator phase. (b)The momentum-resolved defect density  $n(p)$ as a function of the dimensionless momentum $p\sqrt{T/\delta_{max}}$ for the different quench durations $T$ at $\delta_{max}=33$kHz. (c)The dependence of the total defect density $n$ on both the quench range $\delta_{max}$ and the inverse quench rate $\tau_{Q}$. 3D solid red (blue, magenta, green, navy blue, purple) spheres with error bars and their corresponding solid lines represent the experimental data and simulation results for $\delta_{max}=2\pi \times\ 5.25$  ($6.30,7.35,8.40,9.45,10.50$) kHz, respectively. In the side plane, the gray solid (open) circles represent the saturated defect density of the experiment data (simulation). The brown dashed line is the fit of the experimental data to $n\sim \delta^{c}_{max}$. In the bottom plane, the gray solid (open) circles represent the critical inverse quench rate $\tau_{Q,c}$ obtained from the experiment data (simulations). The dashed green line is the fit of the experimental data to $\tau_{Q,c}\sim \delta^{b}_{max}$. }
	\label{pn1}
\end{figure*}

\emph{Experiment with the Rice-Mele model.} To test the universal breakdown of KZM in systems with spatial dimensionality and lattice symmetry, we realize fast quenches in the 1D Rice-Mele model, where the dynamic phase transition occurs from the critical metal phase to the trivial insulator phase, shown in the Fig.~\ref{pn1}(a).  The Hamiltonian in the momentum space is written as
\begin{equation}
\begin{aligned}
H_{RM}(t)=\sum_{k}[(v+w\cos{ka})\sigma_{x}+(w\sin{ka})\sigma_{y}+u(t)\sigma_{z}]
\end{aligned}
\end{equation}
where $u(t)$ is the time-dependent on-site potential. By quenching  $u(t)$ from 0 to a finite value $u_{max}$, we could explore the breakdown of KZM in a topological phase transition. Specifically, we study  the continuum limit $v=w$, and $H_{RM}$ can be mapped to the Hamiltonian $H_{exp}(t)=\sum_{p}(p\sigma_{x}+\delta(t)\sigma_{z})$, following an unitary transformation. 

Experimentally, the phase transition dynamics of the Rice-Mele model is encoded in a trapped ion qubit, which consists of a series of LZ tunnellings with microwave couplings of difference strengths. $\delta(t)=u(t)$ is the time-dependent microwave detuning, and $p=w(ka-\pi)$ is the microwave coupling playing the role of the synthetic Bloch momentum in the lattice model. By quenching $\delta(t)$, $H_{exp}(t)$ describes a phase transition from the critical metal phase to the insulator one~\cite{asboth2016short}.  We take a linear quench scheme with $\delta(t)=vt=\delta_{max} t/T$, and $p\le p_{m}$ with $p_{m}$ is the maximum value of the synthetic Bloch momentum limited by the quench rate. For each $p$, we initialize the system in the lower band $|\Phi \rangle=1/\sqrt{2}(-1,1)^{T}$. Then, we linearly ramp up $\delta(t)$ from 0 to $\delta_{max}$, driving the phase transition in the system. The defect production is quantified as the sum of the net excitations to the upper band $\int_0^{p_m} n(p)dp$, where $n(p)$ is the final population of the upper band. $p_m$ represents the cutoff momentum, which is determined by the size of the nonadiabatic region in momentum space\cite{raeisi2020quench}. For the breakdown of KZM, there exists a critical value of the quench time $T$, which satisfies $T_{c}\delta_{max}\sim 1$, i.e. the critical quench rate $v_{c}=\delta_{max}/T_{c}=\delta^{2}_{max}$, then $p_m=2\pi \sqrt{v}$ for $v<v_c$, while $p_m=2\pi \sqrt{v_c}$ for $v\ge v_c$. 

For the given $\delta_{max}$ and $T$, we drive the evolution of the qubit according to $H_{exp}(t)$ with a sequence of $p$. At the end of evolution, we measure the final density matrix of the qubit through the quantum state tomography. Thus, the momentum-resolved defect density can be obtained by $n(p)=\langle\Psi(T)|\rho(T)|\Psi(T)\rangle$, where $\Psi(T)$ and $\rho(T)$ is respectively the upper eigenstate and the instantaneous density matrix at $t=T$, for a fixed quench range $\delta_{max}$. Fig.~\ref{pn1}(b) shows $n(p)$ for $\delta_{max}=33 kHz$. We define the inverse quench rate as $\tau_Q=T/\delta_{max}$. For the same $T$ we obtain the total defect density by calculating $\int_0^{p_m} n(p)dp$, and the dependence of the total defect density $n$ on $\tau_Q$ is shown by the red dots in Fig.~\ref{pn1}(c). We observe that the total defect density
$n$ is consistent with the power-law behavior predicted by the KZM scaling $n\sim \tau_{Q}^a$ when $\tau_Q \ge T_{c}/\delta_{max}$. From a fit we
obtain $a=-0.51\pm 0.07$, which agrees well with the theoretical prediction $a=-d\nu/(z\nu + 1)=-0.5$ for the equilibrium quantum critical point of Ising universality class with $d=z=\nu=1$\cite{dora2019kibble}. However, for $\tau_{Q}< T_{c}/\delta_{max}$, the total defect density $n$ shows a plateau, which signals the breakdown of the Kibble-Zurek scaling law. 

We further repeat the experiments with varying $\delta_{max}$, of which the resulting defect densities are shown as solid dots with different colors in Fig.~\ref{pn1}(c). For all six demonstrated $\delta_{max}$, the defect densities show two common features. First, they follow the the Kibble-Zurek scaling at large $\tau_Q$, but give a plateau at small $\tau_Q$. The critical value of $\tau_{Q}$ which decreases with increasing $\delta_{max}$, represented by the gray circles in the bottom plane, signifies the breakdown of the Kibble-Zurek scaling. By fitting the critical inverse quench rate $\tau_{Q,c}\sim \delta^{-b}_{max}$, we obtain $b=2.12\pm 0.13$, which agrees well with the theoretical prediction $b=z\nu+1=2$\cite{zeng2023universal}. At small $\tau_Q$, the defect density no longer depends on the quench rate, but grows monotonically with $\delta_{max}$, as shown by the gray circles in the side plane of Fig.~\ref{pn1}(c). From the theoretical prediction of $n\propto\delta^{c}_{max}$, we obtain $c=0.97\pm 0.02$ by fitting, which agrees well with $c=d\nu=1$\cite{zeng2023universal}.

\emph{Conclusion} We have experimentally verified the universal breakdown of the KZ scaling in the fast quenches. Our experimental results shows that this universal behavior emerges not only in the dynamical phase transition of the LZ tunneling, but also in 1D Rice-Mele model realized by a trapped-ion quantum simulator. The breakdown in the respective system can be described by two scaling laws, where the critical inverse quench rate $\tau_{Q,c}\propto \delta_{max}^{-(z\nu+1)}$ and the plateau defect denstiy  $n\propto \delta^{d\nu}_{max}$ is solid verified by the experimental data of the phase transition of a simulator of the 1D Rice-Mele model, as predicted by the universal breakdown mechanism\cite{zeng2023universal}. These observations offer a deep insight for the investigations of fast quenches of quantum phase transitions in many-body systems,  including cold atoms \cite{donadello2016creation,ko2019kibble,goo2021defect,goo2022universal}, 2D Coulomb crystals\cite{guo2024site}, optical quantum gases \cite{ozturk2021observation}, and qubit systems\cite{chertkov2023characterizing,guo2019observation,xu2020probing}. Our studies can also be extended to the open quantum systems, driven by non-Hermitian Hamiltonians, opening new avenues for controlling the quantum phase transitions.

\emph{Acknowledgements} We acklowedges the helpful discussions with Bal{\'a}zs D{\'o}ra and Lei Xiao. This work is supported by The National Key Research and Development Program of China under Grant No.2022YFC2204402, Guangdong Provincial Quantum Science Strategic Initiative under Grant No.GDZX2203001 and No.GDZX2303003, Shenzhen Science and Technology Program under Grant No.JCYJ20220818102003006, the National Natural Science Foundation of China under Grant  No.12275233. 

\bibliography{kz3}

\newpage
\appendix
%

\section*{\textbf{\large Supplementary Materials for\\ ``Verified Universal Breakdown of Kibble-Zurek Scaling in Fast Quenches''}}
\FloatBarrier
\author{Xinxin Rao}
\thanks{These authors contributed equally to this work.}
\affiliation{School of Physics and Astronomy, Sun Yat-Sen University, Zhuhai, 519082, China}

\author{Yang Liu}
\thanks{These authors contributed equally to this work}
\email{liuyang@quantumsc.cn}
\affiliation{Quantum Science Center of Guangdong-Hong Kong-Macao Greater Bay Area, Shenzhen 518045, China}
\affiliation{Guangdong Provincial  Key Laboratory of Quantum Metrology and Sensing, Sun Yat-Sen University, Zhuhai, 519082, China}

\author{Mingshen Li}
\affiliation{School of Science, Sun Yat-sen University, Shenzhen 518100, China}

\author{Teng Liu}
\affiliation{School of Physics and Astronomy, Sun Yat-Sen University, Zhuhai, 519082, China}

\author{Huabi Zeng}
\affiliation{School of Physics and Optoelectronic Engineering,  Hainan University, Haikou, 570228, China}

\author{Le Luo}
\email{luole5@mail.sysu.edu.cn}
\affiliation{School of Physics and Astronomy, Sun Yat-Sen University, Zhuhai, 519082, China}
\affiliation{Quantum Science Center of Guangdong-Hong Kong-Macao Greater Bay Area, Shenzhen 518045, China}
\affiliation{Guangdong Provincial  Key Laboratory of Quantum Metrology and Sensing, Sun Yat-Sen University, Zhuhai, 519082, China}
\affiliation{Shenzhen Research Institute of Sun Yat-Sen University, Shenzhen 518057, China}
\affiliation{State Key Laboratory of Optoelectronic Materials and Technologies, Sun Yat-Sen University, Guangzhou 510275, China}

\date{\today}
%
%
\maketitle

\setcounter{equation}{0}
\setcounter{figure}{0}
\setcounter{table}{0}
\setcounter{page}{1}
\setcounter{section}{0}
\makeatletter
\renewcommand{\theequation}{E\arabic{equation}}
\renewcommand{\thefigure}{S\arabic{figure}}
\renewcommand{\bibnumfmt}[1]{[S#1]}
\renewcommand{\citenumfont}[1]{S#1}
\renewcommand{\thesection}{S\arabic{section}}

\section{Experimental Setup and Quantum Tomography}
\begin{figure}
	\setlength{\abovecaptionskip}{0.cm}
	\setlength{\belowcaptionskip}{-0.cm}
	\includegraphics[width=0.75\linewidth]{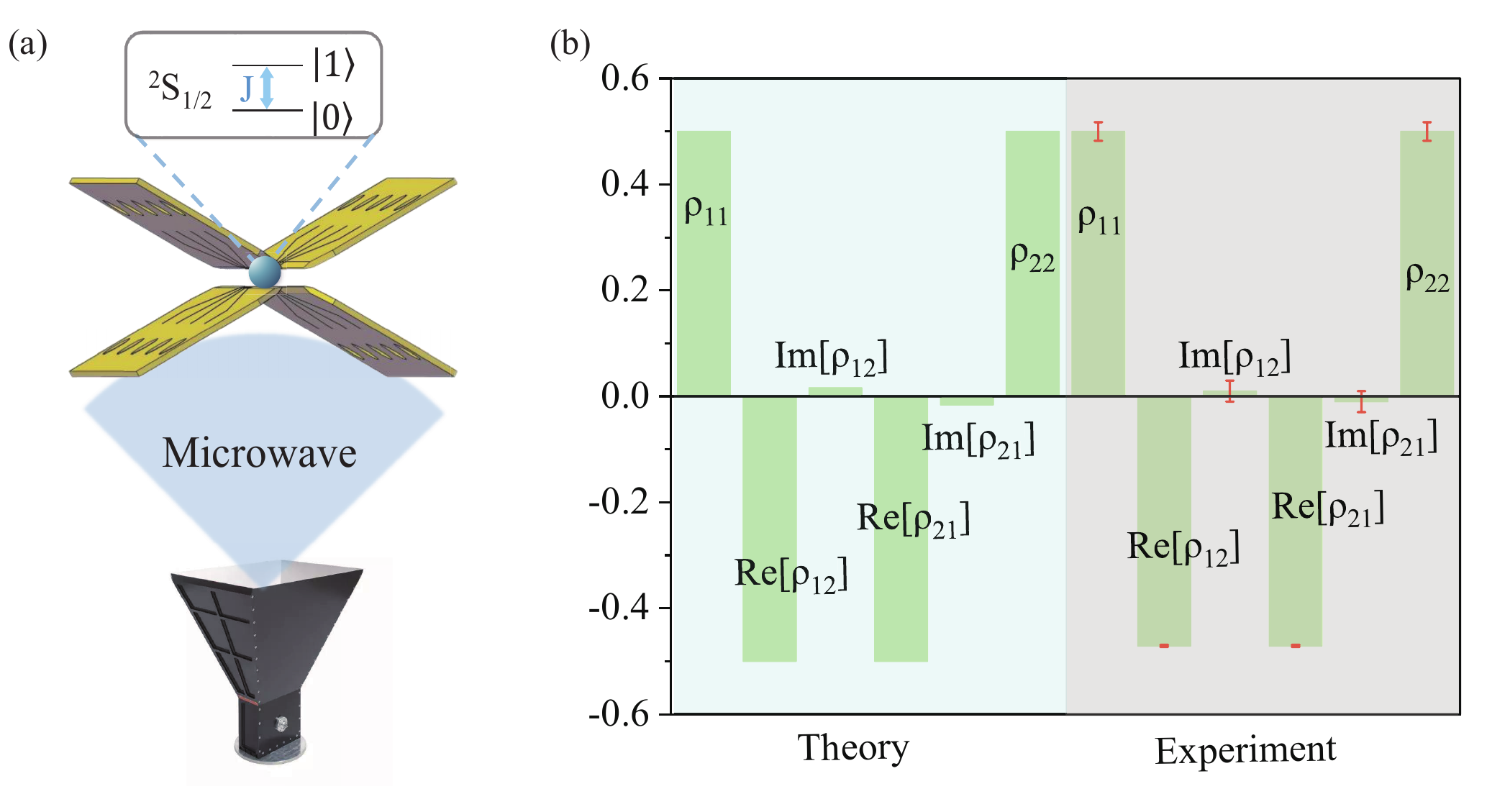}
	\caption{Experiment setup. (a)Schematic diagram of the experimental apparatus; (b) Reconstructed density matrix  for momentum $p=0$ obtained under the conditions shown in Fig.~3(c) of the main text, with $T\delta_{max}=2^{-5}$.}
	\label{rho}
\end{figure}

A single $^{171}$Yb$^+$ ion is confined in a linear Paul trap comprising two sets of radio frequency (RF) and direct current (DC) electrodes constructed from gold-plated ceramic, as shown in Fig.~\ref{rho}(a). A phase-synchronized high-frequency, high-voltage RF signal is applied to the RF electrodes (RF1 and RF2), while a high-precision, low-noise DC voltage is applied to the DC electrodes, generating a three-dimensional trapping potential in the central region of the trap.

The two hyperfine states of the \({}^{2}S_{1/2}\) manifold, \(|F=0, m_F=0\rangle\) and \(|F=1, m_F=0\rangle\), are denoted as \(|0\rangle\) and \(|1\rangle\), respectively. These states maintain a sufficiently long coherence time, preserving the system's quantum properties. A custom-built Helmholtz coil generates magnetic field that defines the quantization axis, lifts the degeneracy of the three Zeeman sublevels, and prevents the ion from being optically pumped into a coherent dark state. For a given field strength \( B \), the energy splitting between the \(|0\rangle\) and \(|1\rangle\) clock states is given by $\Delta E = 12.64281212\,\text{GHz} + 311 B^2\,\text{Hz}$ with $B$ in Gauss. In our experiment, the applied magnetic field is set to approximately \( 8.139 \) Gs.

For precise timing control, the Advanced Real-Time Infrastructure for Quantum Physics (ARTIQ) system orchestrates the cooling, pumping, and detection lasers, along with the microwave signals. The microwave field is generated by mixing signals from an arbitrary waveform generator (AWG) and a standard RF source (Rohde \& Schwarz, SMA 100B). The standard RF source operates at a fixed frequency of $12.611582722GHz$ with a power of $13dBm$, while the quenching process is achieved by dynamically modulating the amplitude and frequency of the AWG output. The AWG runs at $31.25MHz$, corresponding to the resonant transition frequency.

In the experiment, only the population $P_{|1\rangle }$ can be directly measured. However, a general $2\times2$ density matrix contains four real parameters and must satisfy Hermiticity, trace normalization, and positivity constraints. Any such density matrix $\rho$ can be expressed as
\begin{equation}
  \rho = \frac{1}{2} \sum_{i=0}^{3} S_{i}\, \sigma_{i},
  \label{rhodensity}
\end{equation}
where the $S_{i}$ are the Stokes parameters, and $\sigma_{0}$ and $\sigma_{1,2,3}$ denotes the identity and pauli matrices, respectively. In the experimentally define \(|0\rangle\), \(|1\rangle\) basis, the Stokes parameters are determined from projective measurements as
\begin{equation}
	\begin{aligned}
		S_0 &= P_{|0\rangle} + P_{|1\rangle},\\
		S_1 &= P_{\frac{1}{\sqrt{2}}(|0\rangle + |1\rangle)} - P_{\frac{1}{\sqrt{2}}(|0\rangle - |1\rangle)},\\
		S_2 &= P_{\frac{1}{\sqrt{2}}(|0\rangle + i|1\rangle)} - P_{\frac{1}{\sqrt{2}}(|0\rangle - i|1\rangle)},\\
		S_3 &= P_{|0\rangle} - P_{|1\rangle}.
	\end{aligned}
	\label{stokes}
\end{equation}
The population $P_{|0\rangle }$ is inferred by applying a resonant \( \pi \)-pulse to swap the populations of \(|0\rangle\) and \(|1\rangle\). Using this quantum state tomography protocol, the reconstructed density matrix is shown in Fig.~\ref{rho}(b).

\begin{figure}
	\setlength{\abovecaptionskip}{0.cm}
	\setlength{\belowcaptionskip}{-0.cm}
	\includegraphics[width=1.0\linewidth]{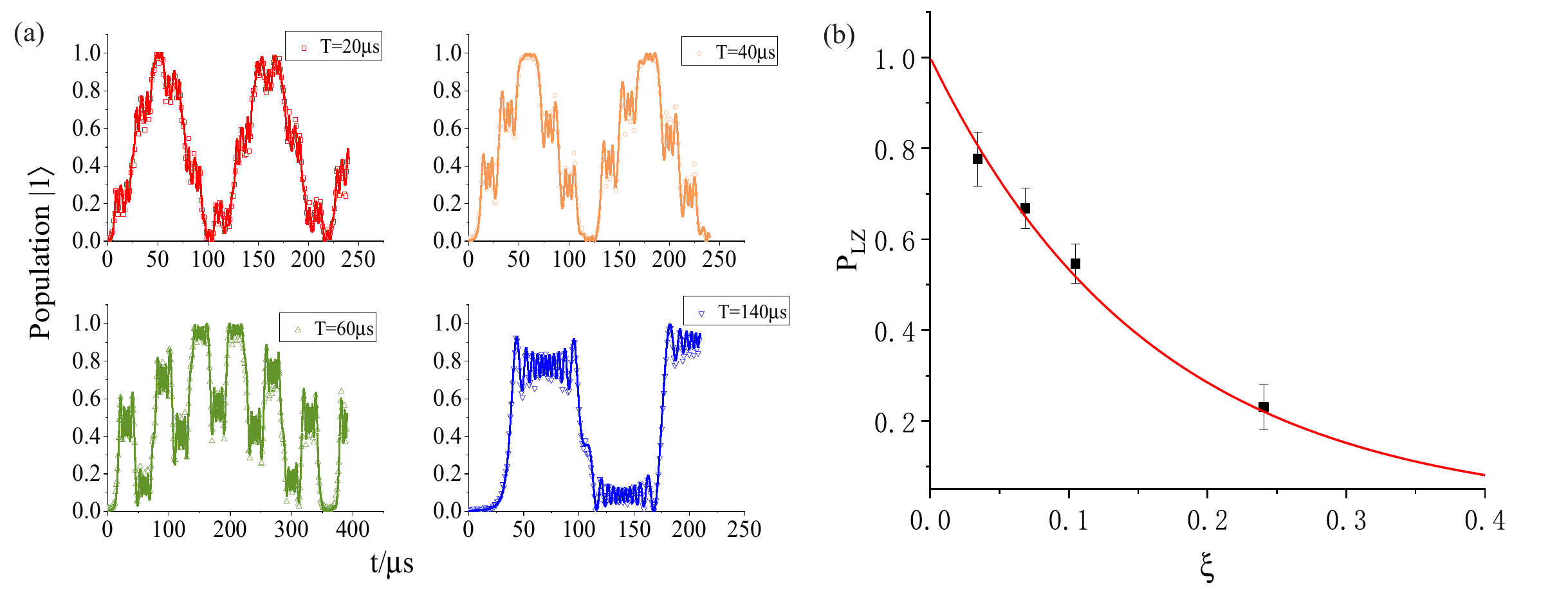}
	\caption{Landau-Zener tunneling dynamics. (a)Time evolution of the projection onto the $|1\rangle$ state under periodic detuning modulation for four different modulation periods; red, $T=20 \mu s$; orange, $T=40 \mu s$; green, $T=60 \mu s$; and blue $T=140 \mu s$. In these measurements,  $\delta_{max}=2\pi \times 300 kHz$ and $J=2\pi\times 18.11kHz$. (b)The measured Landau-Zener tunneling probability with the theoretical prediction. Data points with error bars correspond to the four modulation periods displayed in (a).}
	\label{HermitianLZ}
\end{figure}

\section{Finite-Range Landau-Zener Transition}
Due to the long coherence time, high fidelity of initialization and manipulation and detection, the trapped ion system is an ideal platform for this investigation of the Landau-Zener dynamics\cite{ivakhnenko2023nonadiabatic}. Here, we rewrite the Hamiltonian of the two-level quantum system($\hbar$=1):
\begin{equation}
	H(t)=\begin{pmatrix}
		-\delta(t)/2& J\\
		J & \delta(t)/2
\end{pmatrix},
\label{Heff0}
\end{equation}
where $J$ is the coupling strength and $\delta$ is the detuning. Its eigenvalues are $\pm\frac{1}{2}\sqrt{4J^2+\delta^2}$. The system is initially prepared in an eigenstate of $H(0)$ corresponding to the limit $\delta \gg J $, denoted as $|0\rangle$. The Landau-Zener tunneling and the Landau-Zener-St$\ddot{u}$ckelberg Interferences can be realized by modulating the $\delta$. In our experiments, we apply the triangular waveform modulation. 
\begin{equation}
	\delta(t)=
	\begin{cases} 
		-\delta_{max}+4\frac{\delta_{max}}{T}(t-nT),  \ \ \ \ \ \ \ \ \  t\in[nT,nT+\frac{T}{2}]\\ 
	\delta_{max}-4\frac{\delta_{max}}{T}(t-\frac{T}{2}-nT), \ \ \   t\in[nT+\frac{T}{2},nT+T]. 
	\end{cases}
\end{equation}
where $ n  $ is an integer starting from $0$. The system will pass through the avoided level crossing where $\delta=0$ and the energy gap $\Delta=2J$, and undergoes Landau-Zener tunneling. The population probability will split from $|0\rangle$ into a superposition of $|0\rangle$ and $|1\rangle$, with the probability of the system remaining in $|0\rangle$ described by the well-known Landau-Zener formula $P=\exp{(-2\pi \xi)}$. $\xi=\Delta^{2}/4v$ is the adiabaticity parameter with the sweep velocity $v=\frac{4\delta_{max}}{T}$.

As shown in Fig.~\ref{HermitianLZ}(a), the experimental results are in good agreement with the theoretical simulations under the parameters $\delta_{\text{max}} = 2\pi \times 300~\text{kHz}$ and $J = 2\pi \times 18.11~\text{kHz}$. This condition satisfies the regime $\delta_{\text{max}} \gg J$, where the Landau-Zener formula $P_{\text{LZ}} = \exp(-2\pi \xi)$ becomes applicable. The experimentally measured tunneling probabilities closely follow the theoretical prediction, as shown in Fig.~\ref{HermitianLZ}(b).

\subsection{Symmetric Quench with Finite-Range}
\begin{figure}
	\setlength{\abovecaptionskip}{0.cm}
	\setlength{\belowcaptionskip}{-0.cm}
	\includegraphics[width=1.0\linewidth]{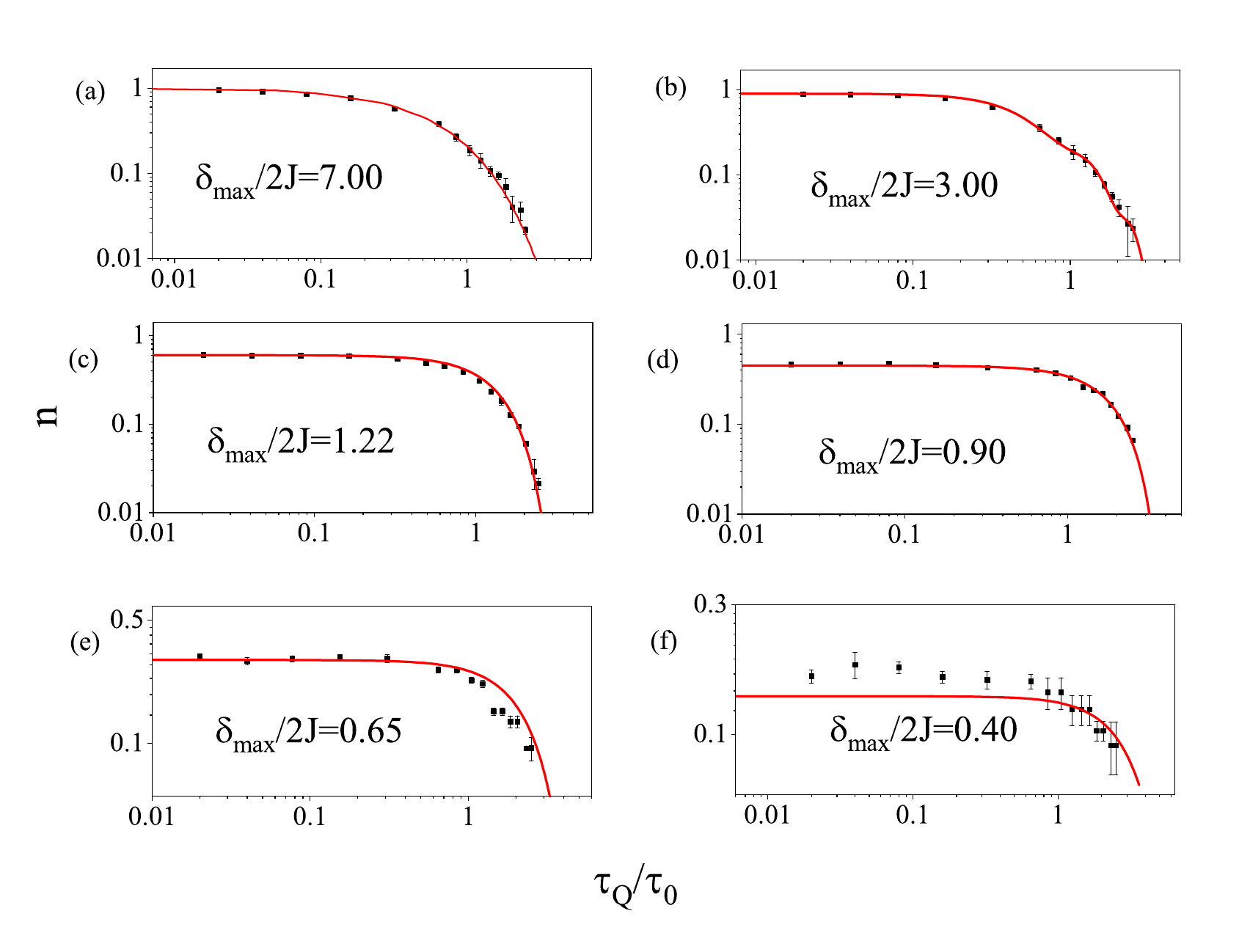}
	\caption{Dynamical phase transition in a two-level Landau-Zener system. The defect density, measured as a function of the inverse quench rate, displays a transition behavior that depends on the quench range. Black dots with error bars are experimental data, and the red curve shows the corresponding theoretical prediction.  Panels (a)--(f) correspond to different quench range $\delta_{max}/2J$: (a) $7.00$, (b) $3.00$, (c) $1.22$, (d) $0.90$,(e) $0.65$, and (f) $0.40$.}
	\label{LZformula}
\end{figure}

For the time-dependent Hamiltonian \ref{Heff0}
with a varying detuning $\delta(t)$ in finite-range(Does not satisgy $\delta \gg J$), we set the eigenvalues ${\lambda}_{\pm}(t) =\pm \sqrt{(\delta(t)/2)^2+J^2}$. Correspondingly, the upper and lower normalized eigenstates are given by $|\Psi(t) \rangle =\{\frac{\lambda-\delta/2}{\sqrt{J^2+(\lambda-\delta/2)^2}},\frac{J}{\sqrt{J^2+(\lambda-\delta/2)^2}}\}^T$ and $|\Phi(t) \rangle= \{\frac{-\lambda-\delta/2}{\sqrt{J^2+(\lambda+\delta/2)^2}},\frac{J}{\sqrt{J^2+(\lambda+\delta/2)^2}}\}^T$, respectively. 

For a given initial state prepared in the upper eigenstate $|\Psi(0) \rangle$, the two detuning setting, $\delta(t)=-\delta_{max}+4\delta_{max}t/T$ for $ t\in[0,T/2] $ and $\delta(t)=4\delta_{max}t/T$ for $ t\in[-T/4,T/4] $, correspond to completely equivalent physical processes. We can obtain the final state after the quench by solving the time-dependent Schr$\ddot{o}$dinger equation $i\frac{\partial}{\partial t} \left | \chi(t) \right \rangle=H(t)\left | \chi(t)  \right \rangle$. we 
plug the ansatz  $| \chi(t) \rangle = \binom{\alpha(t)}{\beta(t)} $ and Hamiltonian \ref{Heff0} to the equation.

\begin{equation}
    \begin{cases} 
     	i\hbar \dot{\alpha(t)} = -\frac{\delta(t)}{2}\alpha(t) + J\beta(t), \\ 
    	i\hbar \dot{\beta(t)} = J\alpha(t) + \frac{\delta(t)}{2}\beta(t). 
    \end{cases}
\end{equation}
By substituting the initial state, we can obtain the final state ($t=T/4$) as:
\begin{equation}
	\begin{cases}
	\alpha(T/4)=\frac{A+B+C}{2\sqrt{2}(E+F)},\\
	\beta(T/4)=-\frac{A^{'}+B^{'}+C^{'}}{E^{'}+F^{'}}
	\end{cases}	
\end{equation}
where $A=2\sqrt{2} \alpha (-\frac{T}{4})D_{-\frac{i*J^2*T}{4\delta_{max}}}[-(-1)^{\frac{1}{4}}(-\frac{T}{4} )\sqrt{\frac{4\delta_{max}}{T}} ]*D_{-1-\frac{i*J^2*T}{4\delta_{max}}}[-(-1)^{\frac{1}{4}} (-\frac{T}{4})\sqrt{\frac{4\delta_{max}}{T}} ] ,\\
 B= 2\sqrt{2} \alpha (-\frac{T}{4})D_{-\frac{i*J^2*T}{4\delta_{max}}}[(-1)^{\frac{1}{4}}(-\frac{T}{4} )\sqrt{\frac{4\delta_{max}}{T}} ]*D_{-1-\frac{i*J^2*T}{4\delta_{max}}}[(-1)^{\frac{1}{4}} (-\frac{T}{4})\sqrt{\frac{4\delta_{max}}{T}} ] ,\\  
 C= (-1)^{\frac{1}{4}}\sqrt{\frac{J^2*T}{\delta_{max}}}*(-\sqrt{2}\beta (-\frac{T}{4})) (D_{-1-\frac{i*J^2*T}{4\delta_{max}}}[-(-1)^{\frac{1}{4}}(-\frac{T}{4} )\sqrt{\frac{4\delta_{max}}{T}} ]^{2}-D_{-1-\frac{i*J^2*T}{4\delta_{max}}}[(-1)^{\frac{1}{4}} (-\frac{T}{4})\sqrt{\frac{4\delta_{max}}{T}} ]^{2}),\\ E=D_{-\frac{i*J^2*T}{4\delta_{max}}}[(-1)^{\frac{1}{4}}(-\frac{T}{4} )\sqrt{\frac{4\delta_{max}}{T}} ]*D_{-1-\frac{i*J^2*T}{4\delta_{max}}}[-(-1)^{\frac{1}{4}} (-\frac{T}{4})\sqrt{\frac{4\delta_{max}}{T}} ],\\
 F=D_{-\frac{i*J^2*T}{4\delta_{max}}}[-(-1)^{\frac{1}{4}}(-\frac{T}{4} )\sqrt{\frac{4\delta_{max}}{T}} ]*D_{-1-\frac{i*J^2*T}{4\delta_{max}}}[(-1)^{\frac{1}{4}} (-\frac{T}{4})\sqrt{\frac{4\delta_{max}}{T}} ],\\
 A^{'}=\sqrt{2} \sqrt{\frac{J^2*T}{\delta_{max}}}\beta (-\frac{T}{4})D_{-\frac{i*J^2*T}{4\delta_{max}}}[-(-1)^{\frac{1}{4}}(-\frac{T}{4} )\sqrt{\frac{4\delta_{max}}{T}} ]*D_{-1-\frac{i*J^2*T}{4\delta_{max}}}[-(-1)^{\frac{1}{4}} (-\frac{T}{4})\sqrt{\frac{4\delta_{max}}{T}} ] ,\\
 B^{'}= \sqrt{2} \sqrt{\frac{J^2*T}{\delta_{max}}}\beta (-\frac{T}{4})D_{-\frac{i*J^2*T}{4\delta_{max}}}[(-1)^{\frac{1}{4}}(-\frac{T}{4} )\sqrt{\frac{4\delta_{max}}{T}} ]*D_{-1-\frac{i*J^2*T}{4\delta_{max}}}[(-1)^{\frac{1}{4}} (-\frac{T}{4})\sqrt{\frac{4\delta_{max}}{T}} ] ,\\  
 C^{'}= 2(-1)^{\frac{3}{4}}*(-\sqrt{2}\alpha (-\frac{T}{4})) (D_{-\frac{i*J^2*T}{4\delta_{max}}}[-(-1)^{\frac{1}{4}}(-\frac{T}{4} )\sqrt{\frac{4\delta_{max}}{T}} ]^{2}-D_{-\frac{i*J^2*T}{4\delta_{max}}}[(-1)^{\frac{1}{4}} (-\frac{T}{4})\sqrt{\frac{4\delta_{max}}{T}} ]^{2}),\\ E^{'}=\sqrt{2}\sqrt{\frac{J^2*T}{\delta_{max}}}D_{-\frac{i*J^2*T}{4\delta_{max}}}[(-1)^{\frac{1}{4}}(-\frac{T}{4} )\sqrt{\frac{4\delta_{max}}{T}} ]*D_{-1-\frac{i*J^2*T}{4\delta_{max}}}[-(-1)^{\frac{1}{4}} (-\frac{T}{4})\sqrt{\frac{4\delta_{max}}{T}} ] , \ and \\
 F^{'}=\sqrt{2}\sqrt{\frac{J^2*T}{\delta_{max}}}D_{-\frac{i*J^2*T}{4\delta_{max}}}[-(-1)^{\frac{1}{4}}(-\frac{T}{4} )\sqrt{\frac{4\delta_{max}}{T}} ]*D_{-1-\frac{i*J^2*T}{4\delta_{max}}}[(-1)^{\frac{1}{4}} (-\frac{T}{4})\sqrt{\frac{4\delta_{max}}{T}} ],	$
where $D_{i*a}[b]$ denotes the parabolic cylinder functions. Then, based on the lower eigenvalue eigenstate of the Hamiltonian at $t=T/4$, which is $|\Phi(T/4) \rangle= \{\frac{-\lambda-\delta_{max}/2}{\sqrt{J^2+(\lambda+\delta_{max}/2)^2}},\frac{J}{\sqrt{J^2+(\lambda+\delta_{max}/2)^2}}\}^T$, the defect density $n=|\langle\Phi(T/4)|\chi(T/4)\rangle|^{2}$ can be calculated for any detuning range. The results are presented as the red line in Fig.~\ref{LZformula}, showing well agreement between theoretical predictions and experimental data.

\subsection{Non-Symmetric Quench with Finite-Range}\label{RiceMele}
\begin{figure}
	\setlength{\abovecaptionskip}{0.cm}
	\setlength{\belowcaptionskip}{-0.cm}
	\includegraphics[width=1.0\linewidth]{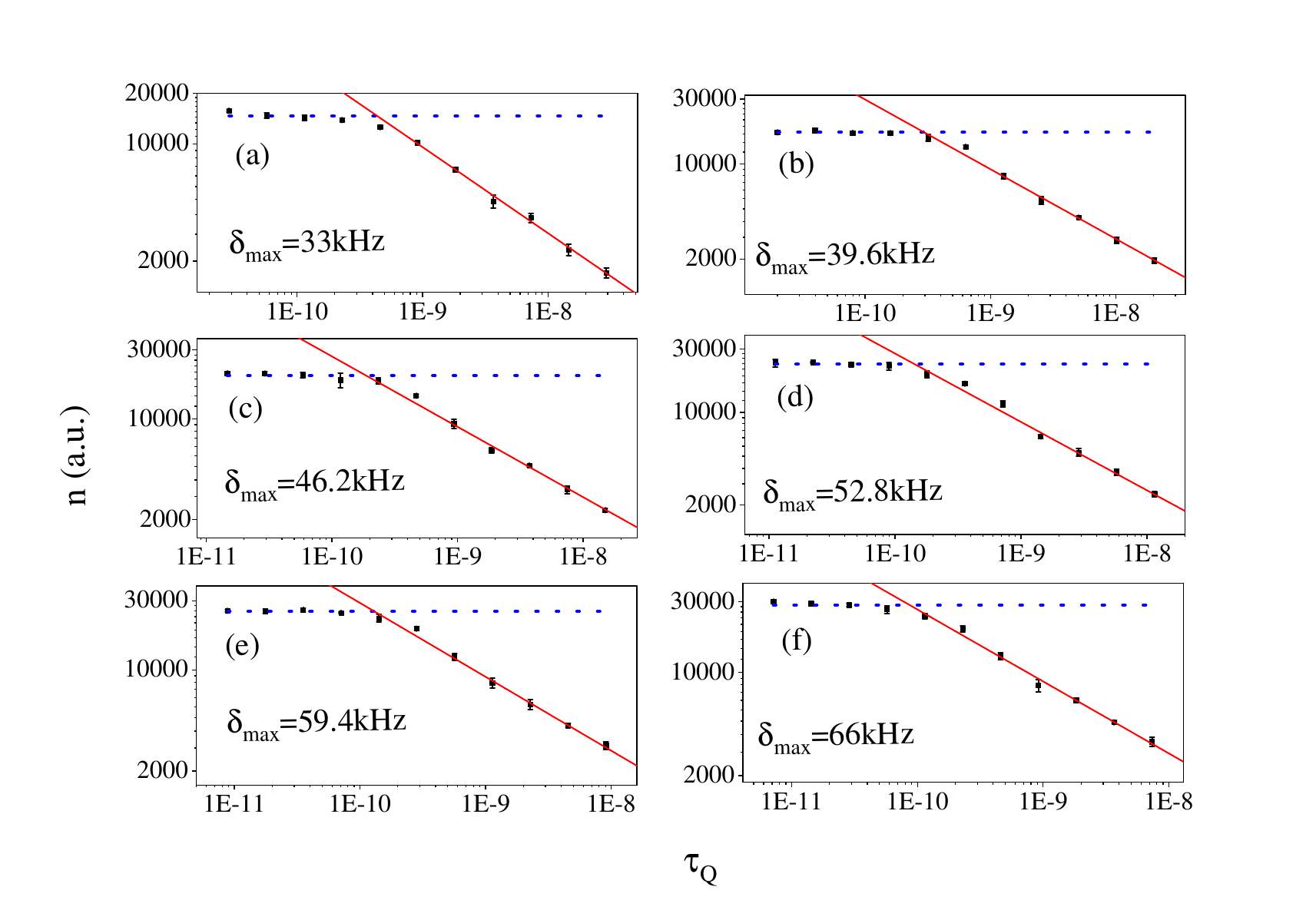}
	\caption{Dynamical phase transition in the Rice-Mele model. In the slow quench regime, the defect density scales with the inverse quench rate as $n\sim \tau_{Q}^a$. The measured power-law exponents are: (a) $ a= -0.513 \pm 0.015$ at $ \delta_{max}= 33$kHz ; (b) $ a=-0.488 \pm 0.024$ at $ \delta_{max}= 39.6$kHz; (c) $ a=-0.510 \pm 0.013$ at $ \delta_{max}= 46.2$kHz ; (d) $ a=-0.512 \pm 0.013$ at $ \delta_{max}= 52.8$kHz; (e) $ a=-0.516 \pm 0.039 $ at $ \delta_{max}= 59.4$kHz; and (f) $ a=-0.487 \pm 0.014 $ at $ \delta_{max}= 66$kHz. These values  agree well with the theoretical prediction of $a=-0.5$.}
	\label{LZformula2}
\end{figure}

When $\delta(t)=\delta_{max} t/T$ for $ t\in[0,T] $, the quench starts at the avoided crossing point, which corresponds to the critical point of the phase transitions. To unify the representation in momentum space, we replace $J$ in the Hamiltonian \ref{Heff0} with $p$. Similar to the symmetric case, assuming the initial state is set as the low-energy eigenstate $|\Phi(0) \rangle =(-1/\sqrt{2}, 1/\sqrt{2})^T$ , the quantum state at any given time can be analytically solved as:
\begin{equation}
	\begin{cases}
		\alpha(T)=A^{''}+B^{''},\\
		\beta(T)=A^{'''}+B^{'''}
	\end{cases}	
\end{equation}
where $A^{''}=\frac{2^{-\frac{5}{2}+\frac{ip^{2}T}{\delta_{max}}}}{\sqrt{\pi}}((-1)^{\frac{1}{4}}\sqrt{\frac{p^2T}{\delta_{max}}}\Gamma[\frac{1}{2}+\frac{ip^{2}T}{8\delta_{max}}]-2\sqrt{2}\Gamma[1+\frac{ip^{2}T}{8\delta_{max}}])*D_{-1-\frac{ip^2T}{4\delta_{max}}}[-(-1)^{\frac{1}{4}}\sqrt{T\delta_{max}} ],\\
B^{''}=-\frac{2^{-\frac{5}{2}+\frac{ip^{2}T}{\delta_{max}}}}{\sqrt{\pi}}((-1)^{\frac{1}{4}}\sqrt{\frac{p^2T}{\delta_{max}}}\Gamma[\frac{1}{2}+\frac{ip^{2}T}{8\delta_{max}}]+2\sqrt{2}\Gamma[1+\frac{ip^{2}T}{8\delta_{max}}])*D_{-1-\frac{ip^2T}{4\delta_{max}}}[(-1)^{\frac{1}{4}}\sqrt{T\delta_{max}} ],\\
A^{'''}=\frac{2^{-\frac{3}{2}+\frac{ip^{2}T}{\delta_{max}}}}{\sqrt{\pi}\sqrt{\frac{p^2T}{\delta_{max}}}}(\sqrt{\frac{p^2T}{\delta_{max}}}\Gamma[\frac{1}{2}+\frac{ip^{2}T}{8\delta_{max}}]-(2-2i)\Gamma[1+\frac{ip^{2}T}{8\delta_{max}}])*D_{-1-\frac{ip^2T}{4\delta_{max}}}[-(-1)^{\frac{1}{4}}\sqrt{T\delta_{max}} ], \ and\\
B^{'''}=\frac{2^{-\frac{3}{2}+\frac{ip^{2}T}{\delta_{max}}}}{\sqrt{\pi}\sqrt{\frac{p^2T}{\delta_{max}}}}(\sqrt{\frac{p^2T}{\delta_{max}}}\Gamma[\frac{1}{2}+\frac{ip^{2}T}{8\delta_{max}}]+(2-2i)\Gamma[1+\frac{ip^{2}T}{8\delta_{max}}])*D_{-1-\frac{ip^2T}{4\delta_{max}}}[(-1)^{\frac{1}{4}}\sqrt{T\delta_{max}} ],$ where $\Gamma$ denotes the Gamma function. Thus, we obtain the analytical solution for the final state of the quench $| \chi(T) \rangle = \binom{\alpha(T)}{\beta(T)} $. 

\begin{figure}
	\setlength{\abovecaptionskip}{0.cm}
	\setlength{\belowcaptionskip}{-0.cm}
	\includegraphics[width=1.0\linewidth]{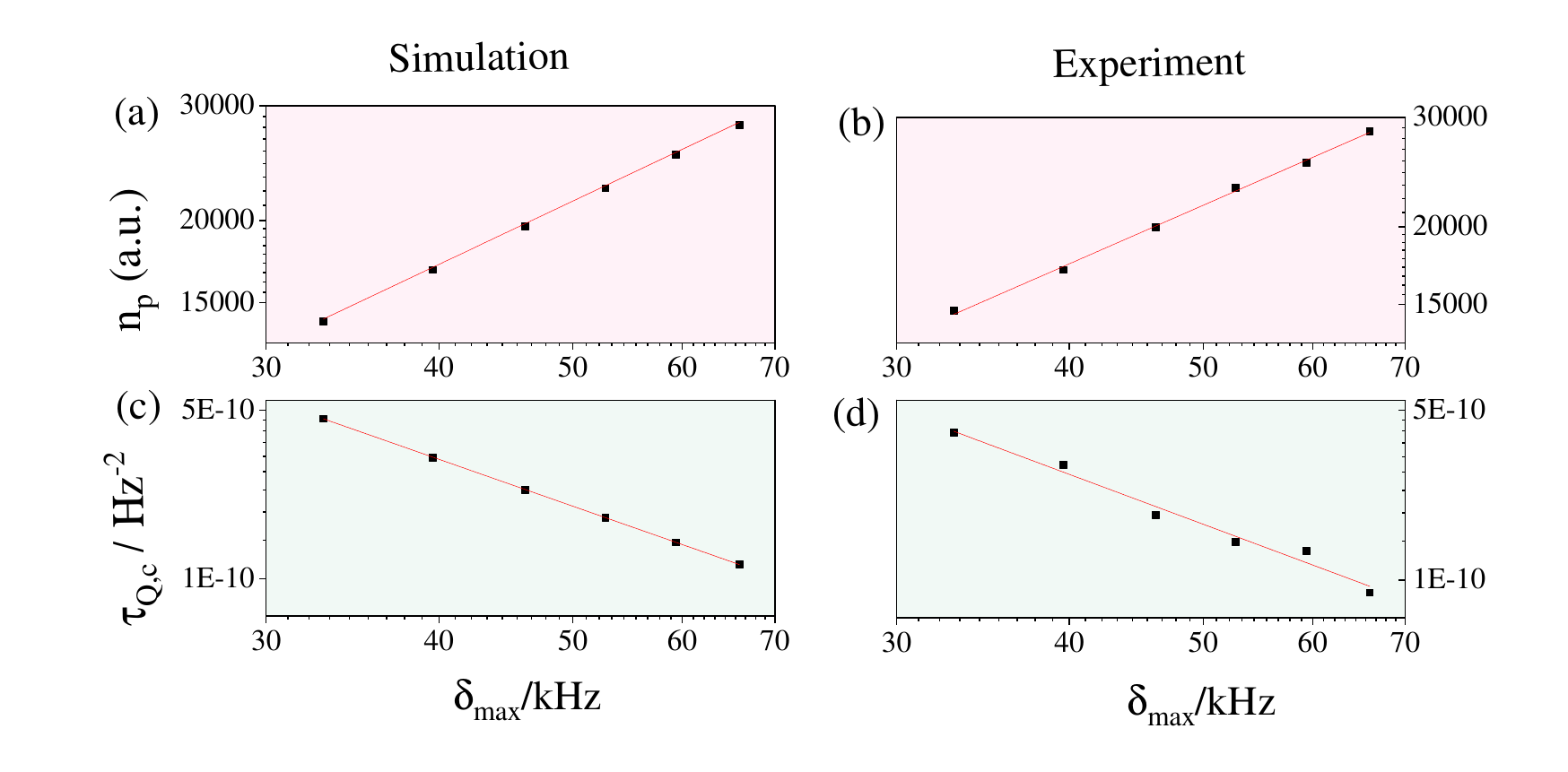}
	\caption{Universal breakdown of Kibble-Zurek scaling in the Rice-Mele model. Panels (a) and (b) show the dependence of the plateau defect density on the quench range. while panels (c) and (d) display the corresponding variation of the critical inverse quench rate with the quench range.}
	\label{LZformula3}
\end{figure}

\section{Defect Density in the Rice-Mele Model}
According to the results in Sec.~\ref{RiceMele}, for a given momentum \( p \),
quench range \([0,\delta_{\max}] \), and quench duration \( T \), the final quantum
state can be expressed as \( | \chi(T) \rangle \). In the RM model, the defect density is determined
by integrating the projection of the time \( T \) quantum state onto the upper eigenstates
of the final Hamiltonian over the entire momentum space. Denoting by
\( |\langle\Psi(T)|\chi(T)\rangle|^2 \) the probability that the state with momentum \( p \) occupies
the upper eigenstate at time \( T \), the total defect density is given by $n = \int  |\langle\Psi(T)|\chi(T)\rangle|^2 \, dp.$ Integration
over all momenta \( p \) simulates defect formation in a one-dimensional Rice--Mele model.
The experimentally measured defect density as a function of the inverse quench rate is shown
in Fig.~\ref{LZformula2}. In the fast-quench regime, the defect density \(n_p=n \) becomes
independent of the quench rate and depends only on the quench range. The dependence
of both \( n_p \) and the critical inverse quench rate \( \tau_Q^c \) on \( \delta_{\max} \) in this
regime is presented in Fig.~\ref{LZformula3}.

All data were independently collected, analyzed, and processed by the authors. All raw data and code are available upon request from the corresponding author.


\end{document}